\documentclass[namedreferences]{solarphysics}

\usepackage[hyperref,optionalrh]{spr-sola-addons} 
\usepackage{graphicx}        
\usepackage{color}           
\usepackage{breakurl}        

\renewcommand{\vec}[1]{{\mathbfit #1}}

\chardef\us=`\_

\begin{document}
\sloppy
\begin{article}
\begin{opening}

\title{Nonaxisymmetric Component of Solar Activity and the Gnevyshev-Ohl rule}

%
\author[addressref={aff1},corref,email={helena@ev13934.spb.edu}]{\inits{E.S.}\fnm{E.S.}\lnm{Vernova}\orcid{0000-0001-8075-1522}}
\author[addressref=aff1]{\inits{M.I.}\fnm{M.I.}~\lnm{Tyasto}}\sep
\author[addressref=aff2]{\inits{D.G.}\fnm{D.G.}~\lnm{Baranov}\orcid{0000-0003-2838-8513}}\sep
\author[addressref=aff1]{\inits{O.A.}\fnm{O.A.}~\lnm{Danilova}}\sep

%
\runningauthor{E.S.~Vernova \textit{et al.}}
\runningtitle{Nonaxisymmetric Component of Solar Activity}

\address[id={aff1}]{IZMIRAN, SPb. Filial, Laboratory of Magnetospheric
Disturbances, St. Petersburg, Russia}
\address[id={aff2}]{Ioffe Institute, St. Petersburg, Russia}

\begin{abstract}
The vector representation of sunspots is used to study the nonaxisymmetric features of the solar activity distribution (sunspot data from
Greenwich--USAF/NOAA, 1874--2016).The vector of the longitudinal asymmetry is defined for each Carrington rotation; its modulus characterizes
the magnitude of the asymmetry, while its phase points to the active longitude. These characteristics are to a large extent free from the
influence of a stochastic component and emphasize the deviations from the axisymmetry. For the sunspot area, the modulus of the vector of the
longitudinal asymmetry changes with the 11-year period; however, in contrast to the solar activity, the amplitudes of the asymmetry cycles obey
a special scheme. Each pair of cycles from 12 to 23 follows in turn the Gnevyshev--Ohl rule (an even solar cycle is lower than the following odd
cycle) or the ``anti-Gnevyshev--Ohl rule'' (an odd solar cycle is lower than the preceding even cycle). This effect is observed in the
longitudinal asymmetry of the whole disk and the southern hemisphere. Possibly, this effect is a manifestation of the 44-year structure in the
activity of the Sun. Northern hemisphere follows the Gnevyshev--Ohl rule in Solar Cycles 12--17, while in Cycles 18--23 the anti-rule is
observed.

Phase of the longitudinal asymmetry vector points to the dominating (active) longitude. Distribution of the phase over the longitude was studied
for two periods of the solar cycle,  ascent-maximum and descent-minimum, separately. The longitudinal distribution displays maximum at the
longitude $\sim 180^\circ$ during ascent-maximum   and at  $\sim 0^\circ/360^\circ$ during descent-minimum. The active longitude changes with
the reversals of polarity of the local (minimum of solar activity) and global magnetic fields (reversal of polar magnetic field).

\end{abstract}
\end{opening}

\section{Introduction}
     \label{S-Introduction}

Observations of the Sun, solar activity and its cyclic changes have a long history, although the beginning of a truly systematic collection of
sunspot data should apparently be attributed to the observations of M. Schwabe and R. Wolf in the mid-19th century. Later it became clear that
variations in activity are connected with changes in the magnetic fields of the Sun. To the first attempts to explain the 11-year cyclicity of
these phenomena belongs the development of the Babcock--Leighton solar dynamo model (\opencite{babck61}; \opencite{leigh69}). This model and a
number of its improvements are based on the notion of axisymmetry of the solar magnetic field (see the survey \opencite{charb10}). However, at
present it is believed that the violation of axisymmetry, which manifests itself in the form of active longitudes, is not accidental, but is one
of important components of the solar cycle. According to (\opencite{ruzm01}), nonaxisymmetric mode even prevails at the time of solar maximum as
compared with the axisymmetric mode. Active longitudes were called 'one of the most interesting manifestations of the solar nonaxisymmetric
magnetic field' (\opencite{pip15}). It was shown that active longitudes appear also on some types of the stars (\opencite{tuom02}) occupying
alternately diametrically opposite longitudes (flip-flop). The problem of active longitudes is treated in a large number of works, both
experimental and theoretical (see, e.g., \opencite{gaiz93}; \opencite{bai03}; \opencite{bigaz04}; \opencite{pip19} and references therein).

Development of solar cycle models which include the generation of nonaxisymmetric magnetic fields should be based on reliable set of data. At
present some of results obtained in different studies of active longitudes are contradictive. E.g., the lifetime of active longitudes according
to different studies varies from several solar rotations (\opencite{detoma00}) to 20--40 \ rotations (\opencite{bumba69}) and even longer than
one 11-year cycle (\opencite{balth83}; \opencite{bai88}; \opencite{jet97}). Around solar minimum, sunspots of the new cycle tend to concentrate
at the same active longitudes as the sunspots of the previous cycle (\opencite{benev99}; \opencite{ben13}).

Results of some studies support the rigid rotation of active longitudes with the nearly Carrington period (\opencite{bumba69});
\opencite{gaiz93}). It was also found in some of theoretical models (\opencite{pip15}) that the dynamo-generated nonaxisymmetric magnetic field
rotates rigidly. The speed of the solar wind and interplanetary magnetic field display a connection with the rigidly rotating active longitudes
(\opencite{neug00}). Other authors, analyzing the experimental data come to a conclusion that the speed of rotation of active longitudes depends
on the latitude and is determined by the Sun's differential rotation speed (\opencite{berd03}). This shows that the active longitude problem is
a topic of current interest and there are many open questions related to this phenomenon.

In the paper \inlinecite{vern04},  the analysis of the distribution of sunspot area over the longitudes showed that active longitudes manifest
themselves successively at $180^\circ$ during the ascend and maximum phases of the solar cycle and at $0^\circ/360^\circ$ during the phases  of
descend and minimum. The change of the location occurs in the period of the sign change of the leading sunspots at the minimum of the solar
activity and in the period of the polar field reversal. An analogous pattern was observed for the X-ray flares of classes M and X
(\opencite{vern05}) and photospheric magnetic fields.

The same localization of active longitudes ($180^\circ$ and $0^\circ/360^\circ$) was discovered in   major solar flares (\opencite{dods68};
\opencite{jet97}). Activity complex connected with a number of solar flares occupied longitudes around $180^\circ$ during solar maximum of 1991
(\opencite{bumba96}). Preferred longitudes shifting from $180^\circ$ to $30^\circ$ were found in CME (Coronal Mass Ejections) distribution
(\opencite{skirg05}) with the change of the location at the time of transition from the ascend-maximum phase to the descend-minimum and vice
versa.

The aim of the present work is to study the nonaxisymmetric component of the solar activity during 13 solar cycles (1874--2016).
Section~\ref{S-Data} describes data used in this study and method of the data treatment. In Section~\ref{S-Asymmetry} evolution of the
nonaxisymmetric component is considered. Section~\ref{S-Phase} is dedicated to the problem of active longitudes. In Section~\ref{S-Conclusions}
main conclusions are drawn.


\section{Data and Method}
\label{S-Data}

To study the nonaxisymmetric component of the solar activity we used the vector representation of the sunspot data. The   areas and
heliocoordinates of sunspots for 1874--2016 were obtained at the site of Royal Observatory, Greenwich--USAF/NOAA Sunspot Data
(https://solarscience.msfc.nasa.gov/greenwch.shtml). A polar vector was assigned to each of the sunspots observed during Carrington rotation,
the magnitude of the vector being set equal to the sunspot area $Sp$ and the phase angle was defined as the heliolongitude of the sunspot $L$.
Individual vectors were combined by vector summing into one resulting vector for each solar rotation (\opencite{vern02}). The magnitude of the
resulting vector could be interpreted as the quantitative characteristics  of the nonaxisymmetric component of solar activity while the phase
angle pointed to the preferred longitude. We shall refer to the resulting vector as the vector of the longitudinal asymmetry and to its modulus
as the longitudinal asymmetry (LA).

\begin{figure}[t]
   \centerline{\includegraphics[width=0.95\textwidth,clip=]{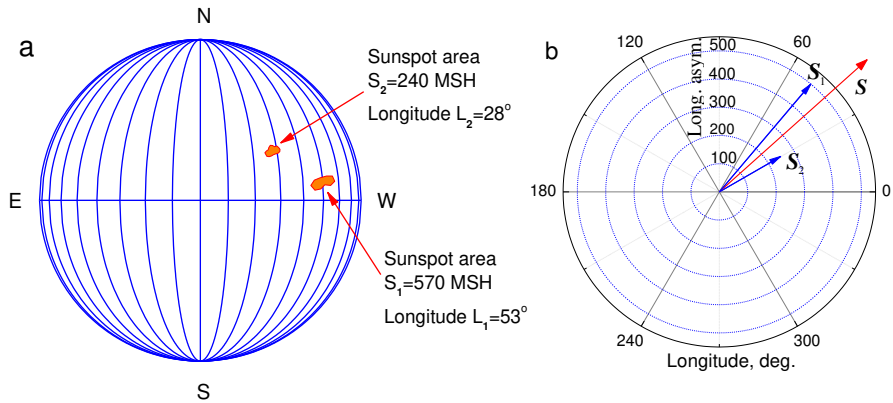}
              }
              \caption{Vector representation of solar activity (scheme).
              (a) Schematic drawing of the Sun with two sunspots on the disk with areas $S_{1}$, $S_{2}$.
              (b) Polar diagram showing two polar vectors assigned to the sunspots 1 and 2.
              The moduli of these vectors are set equal to the sunspot areas, the phases are defined as the longitude  of sunspots.
              The red arrow corresponds to the vector sum of the individual vectors.
                      }
   \label{scheme}
   \end{figure}

The method of the data treatment is illustrated by the scheme in  Figure~\ref{scheme}a,b. Schematic drawing of the  solar disk with two sunspots
is shown in Figure~\ref{scheme}a, while in Figure~\ref{scheme}b the corresponding polar vectors are shown. Resulting vector for the day of
observation is displayed as the red arrow.

For a given day (with number $k$) we calculated the vector sum of all polar vectors $\vec{S}_{i, k}$ corresponding to all $N$ sunspots observed
during this day $(i=1, \dots, N)$. In Figure~\ref{scheme}b the vector sum for two sunspots observed during a day is shown by a red arrow.
Finally, for each solar rotation we defined the resulting vector $\vec{S}$ as the vector sum of daily vectors for all days of the rotation
$(k=1, \dots,27)$. This procedure can be expressed by the equation
\begin{equation}
\label{1} \vec{S} = \sum_{i, k} \vec{S}_{i, k}.
\end{equation}
The vector summing technique allowed us to separate the nonaxisymmetric component of the longitudinal distribution of sunspots represented by
the vector of the longitudinal asymmetry $\vec{S}$, whose magnitude (modulus) will be referred to as the longitudinal asymmetry ($LA=|\vec{S}|$).
The vector phase $L$ points to the dominating longitude.

The vector of the longitudinal asymmetry has the following important features. Sunspot groups living several days or longer make corresponding
number of contributions into the vector of the longitudinal asymmetry $\vec{S}$. Therefore large long-living sunspot groups play the main role
in the magnitude of the longitudinal asymmetry vector. Due to the vector summing  the influence  of the solar activity component uniformly
distributed over longitude was reduced and the stable nonaxisymmetric component of the solar activity distribution was singled out.

Due to the vector summing method the influence of the stochastic component of the solar activity uniformly distributed over longitude was
reduced and the stable nonaxisymmetric component of the solar activity distribution was singled out.

The vector of the longitudinal asymmetry was defined for an integer number of days ($T=27$ days), while a Carrington rotation period is 27.2753
days. The time shift which occurs as a result of this difference is not large but it accumulates and can affect significantly the assignment of
the longitudinal asymmetry to a specific time. To compensate this time shift we used the method where for several rotations the asymmetry was
calculated for the period $T=27$ days, after which for one rotation the asymmetry was calculated for the period $T=28$ days. These scheme
allowed to avoid accumulation of errors and to get the time shift of at most 5 days during the whole, almost 150 year long, period under
consideration.

The longitudinal asymmetry vector can be also computed for some specific groups of sunspots, e.g., for sunspots whose area is above (or below)
some chosen threshold.


\section{Nonaxisymmetric component of the solar activity (modulus of the vector of the longitudinal asymmetry)}
\label{S-Asymmetry}

\subsection{Time variation of the longitudinal asymmetry}
\label{S-Time} Using the method described above we calculated the magnitude and phase angle of the vector of the longitudinal asymmetry for the
whole disc of the Sun as well as separately for the northern and southern hemispheres.  The modulus of the vector of the longitudinal asymmetry
(to be short, the longitudinal asymmetry, or LA)  can be considered as a quantitative characterstics of the nonaxisymmetric component of the
solar activity.

\begin{figure}
   \centerline{\includegraphics[width=0.85\textwidth,clip=]{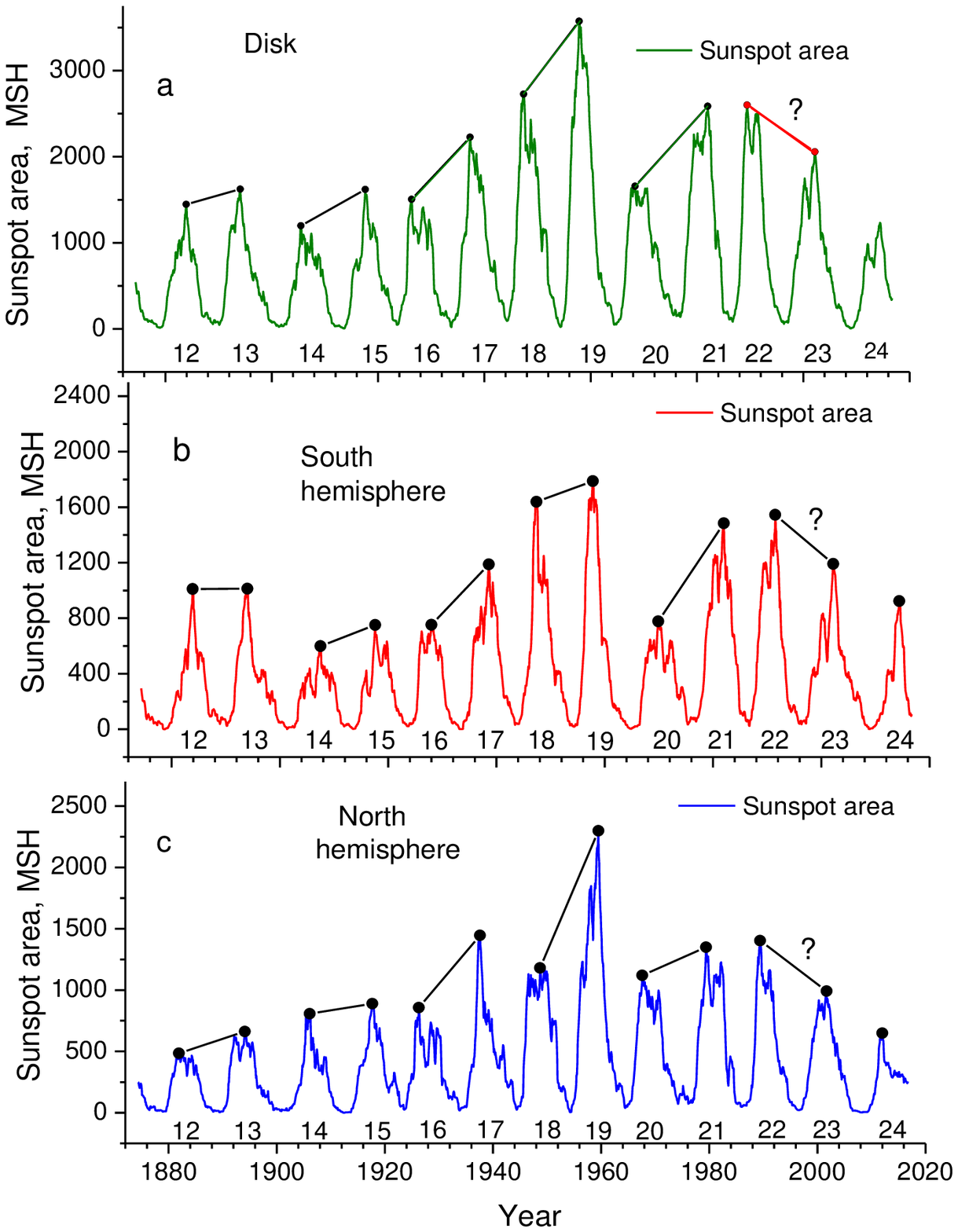}
              }
              \caption{Sunspot area for Solar Cycles 12--24 (1874--2016): (a) solar disk; (b) southern hemisphere; (c) northern hemisphere.
              Solar Cycle numbers are shown at the bottom of the panels.  Maximums of the even cycles ($2N$) and the following odd cycles ($2N+1$)
              are connected by straight line segments to emphasize the manifestation of the Gnevyshev--Ohl rule. Curves were smoothed
              using 13-rotation running mean.
              }
   \label{area}
   \end{figure}
We compared the  variations of the longitudinal asymmetry during 13 solar cycles with the corresponding variations of the sunspot area.In
Figure~\ref{area} changes in time (1874--2016) of the sunspot area are shown for the whole disc (Figure~\ref{area}a), southern hemisphere
(Figure~\ref{area}b) and northern hemisphere (Figure~\ref{area}c).

One of the repeating patterns of the solar activity variations is the Gnevyshev--Ohl rule (\opencite{gnev48}). There exist different
formulations of this rule which manifests itself in many solar activity parameters. The Gnevyshev--Ohl rule as well as its violations are
treated in numerous papers (e.g., \opencite{nagov09}, \opencite{ogur12}, \opencite{java13}).

One of the forms of the Gnevyshev--Ohl (G-O) rule is as follows: the amplitude of an even solar cycle (cycle $2N$) is lower than the amplitude
of the following odd cycle (cycle $2N+1$). The Gnevyshev--Ohl rule holds for ten solar cycles (Solar Cycles 12--21). This pattern is shown in
Figure~\ref{area}; to stress the relation between the cycle heights the straight line segments are drawn connecting the maxima of an even and of
the consequent odd cycles. It is seen that the G-O rule holds similarly for the disc and for the southern and the northern hemispheres. It
should be noted that in the sunspot area the highest is Solar Cycle 19 and one of the lowest is Solar Cycle 20, and these features are present
for the whole disc as well as for the both hemispheres.

It was shown in many studies that the G-O rule does not hold for the pair of Solar Cycles 22--23. This violation of the general rule can be seen
in Figures~\ref{area}a,b,c also. As an explanation of the G-O rule violation in Solar Cycles 22--23 \inlinecite{java16} points out to the
contribution of small sunspot groups with area below 100 MSH (millionths of solar hemisphere). The number of such groups was large in the
southern hemisphere during Solar Cycle 22, whereas in Solar Cycle 23 the number of such groups was low in both hemispheres

\begin{figure}
   \centerline{\includegraphics[width=0.85\textwidth,clip=]{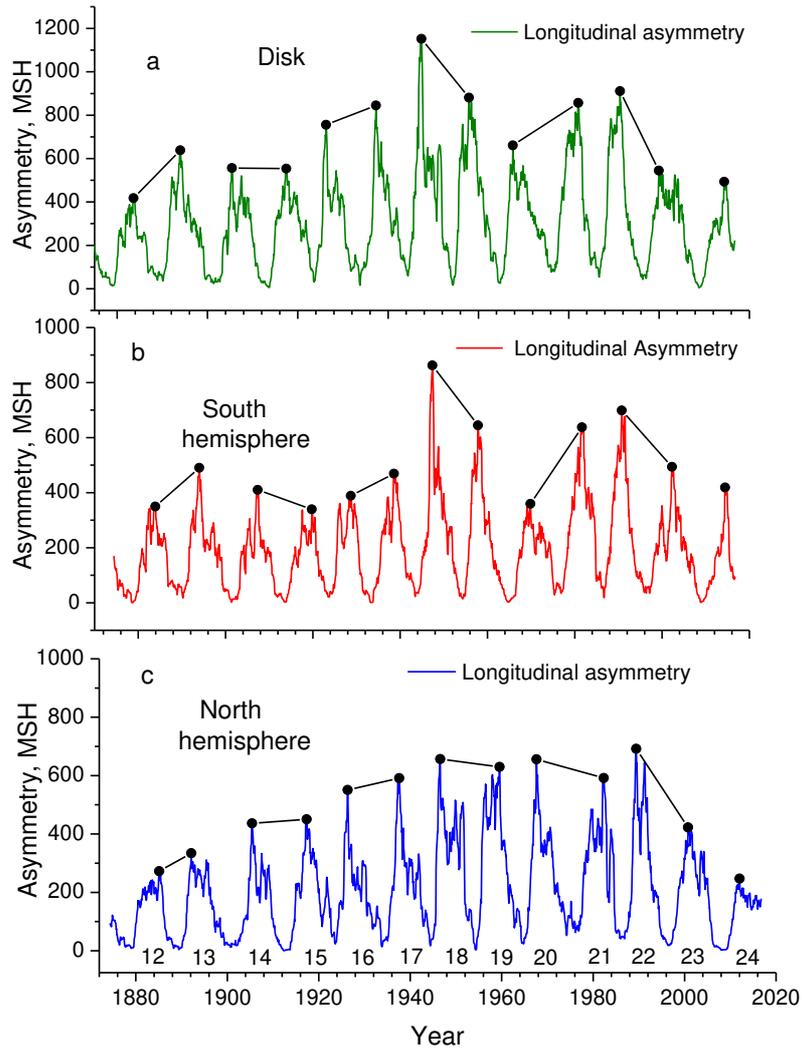}
              }
              \caption{Modulus of the longitudinal asymmetry vector for Solar Cycles 12--24 (1874--2016):
              (a) solar disk; (b) southern hemisphere; (c) northern hemisphere.  Cycle numbers are shown
              at the bottom of the figure.  Maxima of the even cycles ($2N$) and the following odd cycles ($2N+1$)
              are connected with straight line segments to emphasize the alternation of the cycle pairs following
              the Gnevyshev--Ohl rule or anti-Gnevyshev--Ohl rule. Curves were smoothed using 13-rotation running mean.
                      }
   \label{lonasym}
   \end{figure}

Time variations of the nonaxisymmetric component of the solar activity  are shown in Figure~\ref{lonasym} for Solar Cycles 12--24 for the disc
(Figure~\ref{lonasym}a), southern hemisphere (Figure~\ref{lonasym}b) and northern hemisphere (Figure~\ref{lonasym}c).

We use this specific order of the figures (disc, southern hemisphere, northern hemisphere)
to emphasize the similarity of the dependencies for the disc and for the southern hemisphere.
On the other hand, the modulus of the vector of the longitudinal asymmetry for the northern
hemisphere differs considerably from the whole disc and from the southern hemisphere.

A common feature in the variation of the sunspot area (Figure~\ref{area}) and of the longitudinal asymmetry (Figure~\ref{lonasym}) is that they
change in phase following the 11-year solar cycle. At the same time, a comparison of these two parameters show significant differences between
them. The longitudinal asymmetry vector is 3-4 times smaller than the sunspot area due to reduced contribution of the stochastic component
uniformly distributed over longitude. In the course of the cyclic changes the values of the LA do not always follow the changes of the sunspot
area. Thus, Cycle 19 is the highest for the solar activity (Figure~\ref{area}), while for the LA the highest values are attained in Cycle 18 for
the disc and for the southern hemisphere and in Cycle 22 for the northern hemisphere (Figure~\ref{lonasym}). In Cycle 20 the sunspot area was
low for the disc, the southern and the northern hemispheres, whereas this cycle has one of the highest values of the LA for the northern
hemisphere.

 This difference between  the amplitudes of the sunspot area cycles  (Figure~\ref{area}) and of the  LA cycles
 (Figure~\ref{lonasym}) leads to the transformation of the Gnevyshev--Ohl rule in the latter case. The relation between maxima of the
 two successive cycles (the even and following odd cycles) is marked in Figure~\ref{lonasym} by the line segments.

Longitudinal asymmetry of the sunspot distribution shows the following structure. Only the first pair of each four cycles followed the
Gnevyshev--Ohl rule (the amplitude of an even solar cycle  is lower than the amplitude of the following odd cycle); the next pair followed the
``anti-Gnevyshev--Ohl rule'', i.e. the even cycle (cycle $2N$) exceeded the following odd cycle (cycle $2N+1$). This is true for Solar Cycles
12--23. Possibly, this pattern is a manifestation of the 44-year periodicity of the solar activity or the double Hale cycle.

It should be noted that the 44-year pattern described above was observed for the whole solar disc (Figure~\ref{lonasym}a) and was especially
pronounced in the southern hemisphere (Figure~\ref{lonasym}b). In the northern hemisphere (Figure~\ref{lonasym}c) for the modulus of the
longitudinal asymmetry an essentially different pattern is seen for 12 solar cycles (Cycles 12--23). During the first 6 cycles (Cycles 12--13,
14--15, 16--17) the Gnevyshev--Ohl rule holds -- an even solar cycle was lower than the following odd cycle. However, the next 6 cycles followed
the anti-Gnevyshev--Ohl rule -- an even cycle was higher than the following odd cycle. In the northern hemisphere (Figure~\ref{lonasym}c) the
amplitude of the longitudinal asymmetry sharply decreased in the beginning and at the end of time interval under consideration (low values of
the modulus of the vector of the LA in Cycles 12, 13, 23 and 24).

\begin{figure}
   \centerline{\includegraphics[width=0.95\textwidth,clip=]{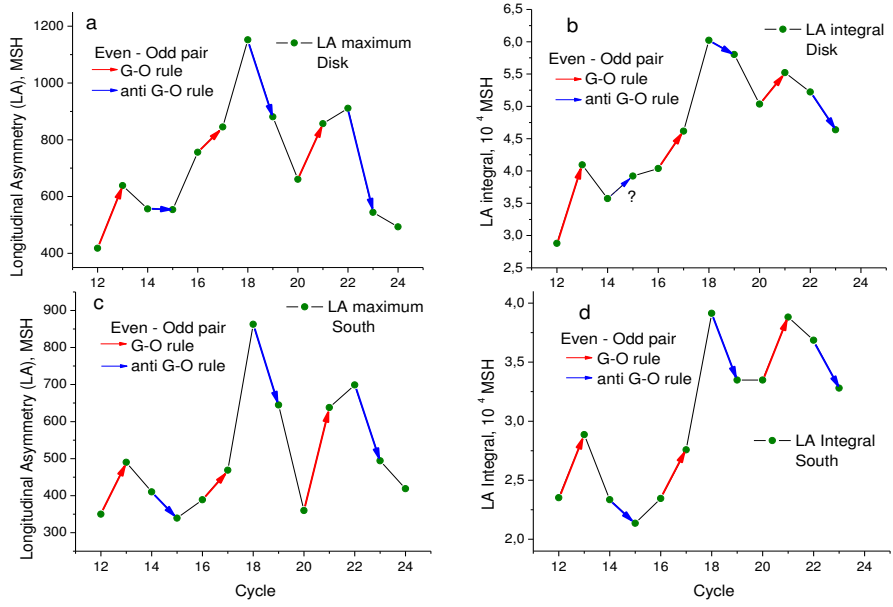}
              }
              \caption{ (a,c) Maxima of the longitudinal asymmetry in Solar Cycles 12--24: (a) solar disk;
              (c)  southern hemisphere. (b,d) Integral of the longitudinal asymmetry for individual solar cycles:
              (b)  solar disk; (d) southern hemisphere. The red arrows show the pairs of even--odd cycles
              following the G-O rule. The blue arrows show the pairs of even--odd cycles following anti-G-O rule.
              In all of the cases (except one) the even-odd pairs show in turn the G-O rule or anti-G-O rule.
                      }
   \label{maxint}
   \end{figure}
\begin{figure}
   \centerline{\includegraphics[width=0.95\textwidth,clip=]{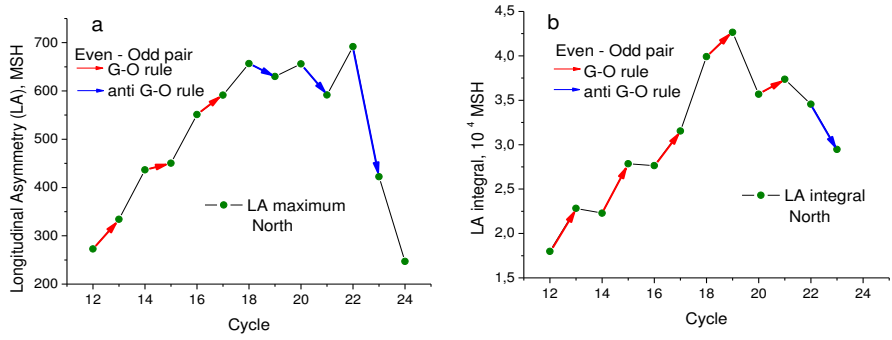}
              }
              \caption{ Northern hemisphere in Solar Cycles 12--24: (a) Maximum of the longitudinal asymmetry.
              (b)  Integral of the longitudinal asymmetry for individual solar cycles. The red arrows show the pairs
              of even--odd cycles following G-O rule. The blue arrows show the pairs of even--odd cycles following
              anti-G-O rule. In contrast to the disk and the southern hemisphere, the maxima of the longitudinal
              asymmetry in the northern hemisphere follow G-O rule only for Solar Cycles 12--17, while for Solar
              Cycles 18--23 anti-G-O rule holds. Integral of the longitudinal asymmetry changes in the same way
              as the area of sunspots (see Figure~\ref{area}2).
                      }
   \label{north}
   \end{figure}


\subsection{Maximum and integral of the longitudinal asymmetry}
\label{S-Integral} To show more clearly the alternation of high and low cycles, we considered the maxima of the longitudinal asymmetry for each
of the cycles. Figure~\ref{maxint} shows the maximum of the LA values for the disk (Figure~\ref{maxint}a) and the southern hemisphere
(Figure~\ref{maxint}c). To compare the heights of adjacent cycles in Figure~\ref{maxint} straight line segments were drawn connecting the maxima
of even and subsequent odd solar cycles: the cases where the G-O rule holds are shown in red, the cases of the anti-G-O rule are marked by blue.
It can be clearly seen that for the disc (Figure~\ref{maxint}a) and the southern hemisphere (Figure~\ref{maxint}c) there is an alternation of
the G-O rule and the anti-G-O rule. These figures illustrate the main features of the variation of the modulus of the longitudinal asymmetry
noted in the analysis of Figure~\ref{lonasym}. An interesting feature is the good correlation (correlation coefficient $R = 0.86$) and an almost
complete matching of the form of the dependencies for the disk (Figure~\ref{maxint}a) and for the southern hemisphere (Figure~\ref{maxint}c),
for which the highest cycle of longitudinal asymmetry was Solar Cycle 18 and the second-largest cycle was Solar Cycle 22. For the solar disk and
for the southern hemisphere, Solar Cycles 12, 15, 20 and 24 have the lowest height of the LA.

Similar to the research of \inlinecite{gnev48}, where the integral characteristic of the solar cycle was used (the sum of Wolf numbers for each
cycle), we considered the integral of longitudinal asymmetry calculated for each solar cycle. We considered the manifestation of the G-O rule
for the longitudinal asymmetry integral whose variation in Solar Cycles 12--24 is shown in Figure~\ref{maxint}b (disc) and Figure~\ref{maxint}d
(southern hemisphere).

The longitudinal asymmetry integral most clearly shows the alternation of the G-O rule and anti-G-O rule for the southern hemisphere
(Figure~\ref{maxint}d.) There is one exception (the pair of Solar Cycles 14--15) for the disk.

It can be seen that the integral for the disc and for the southern hemisphere change in a similar manner, reaching the maximum in Solar Cycle 18
and the second maximum in Solar Cycle 21. On the whole, all four dependencies of Figure~\ref{maxint} resemble each other. The pair of cycles in
which the G-O rule holds and the next pair obeying the anti-G-O rule form a stable pattern that repeats three times over the period in question.

\begin{figure}
   \centerline{\includegraphics[width=0.95\textwidth,clip=]{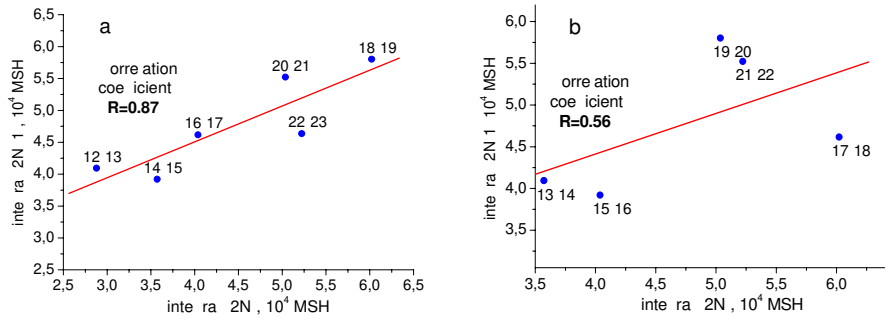}
              }
              \caption{ Correlation of the longitudinal asymmetry maxima for the pairs of cycles: (a) an even cycle ($2N$) vs the following odd
              cycle ($2N+1$); correlation coefficient $R= 0.87$.
(b) an even cycle ($2N$) vs the previous odd cycle ($2N-1$); correlation coefficient $R= 0.56$.
                      }
   \label{corr}
   \end{figure}
For the northern hemisphere, which behaves in a quite different way when compared to the disk and the southern hemisphere, the variation of the
maximum and of the integral of longitudinal asymmetry in Solar Cycles 12--23 is shown in Figure~\ref{north}. In this case, the cycle maximum for
longitudinal asymmetry follows the G-O rule for the first three cycle pairs, and for the next three cycle pairs the anti-G-O rule holds
(Figure~\ref{north}a). High values of asymmetry maximum were observed in 6 consecutive cycles starting from Solar Cycle 17 to Solar Cycle 22,
with the highest value of asymmetry in Solar Cycle 22. On the other hand, the integral of the longitudinal asymmetry of the northern hemisphere
(Figure~\ref{north}b) follows the G-O rule in the same way as the area of sunspots (Figure~\ref{area}). As in the case of the  sunspot area, the
highest cycle is Solar Cycle 19. Thus, the integral of the LA for the northern hemisphere shows no distinctions from the usual sunspot area
variation.

There exists the following formulation of the G-O rule (\opencite{gnev48}): the total Wolf number per even cycle (cycle $2N$) correlates well
with the subsequent odd cycle (cycle $2N+1$), forming a pair of even-odd cycle (correlation coefficient $R = 0.91$). In contrast, the
correlation of the even cycle (cycle $2N$) to the previous odd cycle (cycle $2N-1$) is weak ($R = 0.5$). We verified whether this phenomenon
takes place for the longitudinal asymmetry integral. Figures~\ref{corr}a,b illustrate two types of relations of the LA integral: for even-odd
(a) and for odd-even cycles (b). In the first case, the correlation coefficient equals $R = 0.86$, while in the second case $R = 0.56$. These
values are close to those obtained in (\opencite{gnev48}). It should be emphasized that the coincidence of the results was obtained despite the
fact that essentially different parameters were considered (the sum of Wolf numbers per cycle and the integral of longitudinal asymmetry), as
well as different time intervals (1700--1945 and 1874--2016, respectively). This supports the conclusion of \inlinecite{gnev48} that a pair of
even and subsequent odd cycles form a joint structure.

\begin{figure}
   \centerline{\includegraphics[width=0.95\textwidth,clip=]{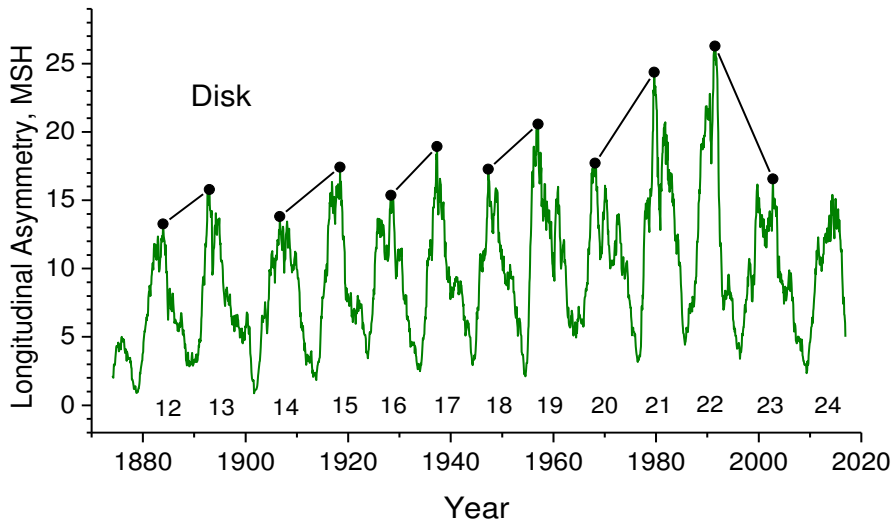}
              }
              \caption{ Small sunspots (area $Sp < 50$ MSH). Modulus of the longitudinal asymmetry vector for the solar disk
              (1874--2016) changes in the same way as the sunspot area (see Figure 2). Curves were smoothed using 13-rotation running mean.
                      }
   \label{small}
   \end{figure}

It is known that large and small sunspots form two populations with different properties. E.g., it was shown in \inlinecite{nagov18} that
sunspots with area $Sp>50$ and $Sp<50$ MSH have different times of life and different latitude distributions. In Figure~\ref{lonasym}, the
changes of the longitudinal asymmetry for the sunspots of arbitrary area were presented. The longitudinal asymmetry for the sunspots with area
smaller than 50 MSH is shown in Figure~\ref{small}. It is seen that the LA of small sunspots obeys the usual G-O rule, as well as the sunspot
area does (Figure~\ref{area}). Thus, the features of the longitudinal asymmetry which manifest themselves in alternation of G-O rule and
anti-G-O rule and the corresponding 44-year structure are characteristic for long-living sunspots with large area.

The presence of a 44-year periodicity was observed in various parameters of the solar activity. For example, regular drops of the equatorial
rotation rate of the Sun were found (\opencite{java03}) with a time gap of 44 years. This effect was interpreted as a manifestation of a double
Hale cycle. The analysis of the large sunspot group diameters (\opencite{efim18}) also shows the 44-year period in the variations of solar
activity. There are some indications that the 44-year periodicity of solar activity is reflected in cosmic ray intensity. The study of
connection between the solar activity and cosmic ray modulation showed that dimensions of the modulation region change not only with 11 and 22
year periods, but also with a period of 44 years (\opencite{dorman99}).

\section{Active longitudes (phase of the longitudinal asymmetry vector)}
\label{S-Phase} 
\begin{figure}
   \centerline{\includegraphics[width=0.45\textwidth,clip=]{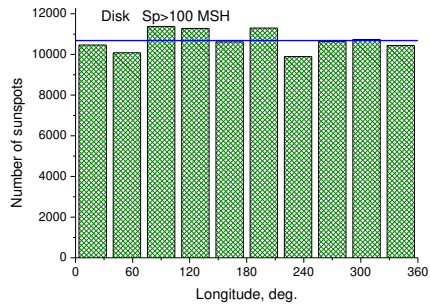}
              }
              \caption{ Longitudinal distribution of the sunspot area ($Sp > 100$ MSH) for 13 solar cycles (1874--2016).  No preferred
              longitudes can be seen.
                      }
   \label{nopref}
   \end{figure}

Asymmetry of the distribution of the solar activity  over the heliolongitude manifests itself in the form of the so-called
active or preferred longitudes. In this section the extension of our previous works on the problem of active longitudes (\opencite{vern04};
\opencite{vern05}; \opencite{vern07}) is presented. Only large sunspots ($Sp>100$ MSH) were considered in this section. The longitudinal
distribution of sunspots for 13 solar cycles (1874--2016) is shown in  Figure~\ref{nopref}. The distribution over longitude is almost uniform
and gives no evidence for the presence of preferred longitudes. However, the picture changes drastically if we study different periods of the
solar cycle separately.  In (\opencite{vern04}) it was found that the distributions of sunspots over longitude have similar forms for the
periods of the ascent and maximum on one hand and for the descent and minimum on the other hand. Therefore, in our analysis these parts of the
cycle were considered jointly as the ascent-maximum epoch (AM) and the descent-minimum epoch (DM).

The two characteristic periods (AM and DM) are connected with the magnetic cycle of the Sun. The boundaries of these characteristic periods
coincide with the change of the polar field sign and the change of the leading sunspot sign in a hemisphere. Thus AM period corresponds to the
interval from the solar minimum to the polar field reversal and DM -- to the interval from the  reversal to the  minimum.

\begin{figure}
   \centerline{\includegraphics[width=0.95\textwidth,clip=]{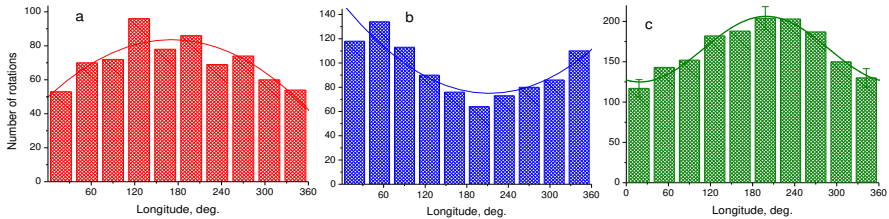}
              }
              \caption{ Distribution of the phase of the longitudinal asymmetry vector for 13 solar cycles (1874--2016).
              (a) AM phase  of the solar cycles. (b) DM phase; maximum of the distribution moves by $180^\circ$ during the
              transition from AM to DM phase and vice versa (flip-flop effect). (c) Joint distribution for AM and DM phases
              obtained by summing the AM histogram with the DM histogram shifted by $180^\circ$. Enveloping curve -- approximation
              with a 2nd degree polynomial.
                      }
   \label{phase}
   \end{figure}

As stated above, the vector summing of the sunspot data allows us to separate the nonaxisymmetric component from the axisymmetric part of solar
activity, which is distributed uniformly over the longitude. This method proved to be useful when studying the time variation of the
longitudinal asymmetry (Section~\ref{S-Asymmetry}). For each Carrington rotation the phase of the longitudinal asymmetry vector points to the
active longitude where large long-living sunspots were observed. If the vector of the longitudinal asymmetry has small magnitude  then the value
of the phase angle becomes unreliable. Therefore while investigating the longitudinal asymmetry we took into account only those cases where the
modulus of the vector exceeded 20 MHS.

Distributions of the phase angle over the longitude for 1874--2016 are plotted in Figure~\ref{phase} for two periods, AM and DM, separately.
These histograms  display two forms of the enveloping curves: convex or concave. Corresponding maxima of distribution are located at longitudes
$\sim180^\circ$ for ascent-maximum (a) and $\sim 0^\circ/360^\circ$ for descent-minimum (b). Comparison of Figures~\ref{phase}a and
Figures~\ref{phase}b shows that there are two similar distributions with the shift of $180^\circ$ with respect to each other. We combined both
histograms plotting the sum of histogram (a) with the histogram (b) shifted by $180^\circ$. Resulting histogram (Figure~\ref{phase}c) includes
data of the whole interval under study (1874--2016). As an enveloping curve for the resulting histogram a sinusoidal function of the following
form was chosen:

\begin{equation} \label{Eq-Hist}
y=A\sin((x-B)/w)+C
\end{equation}

Clearly displayed maximum of the envelope supports the idea of the active longitude jumping by $180^\circ$ (so-called flip-flop) at
the boundaries between AM and DM periods. Stability of the longitude asymmetry maximum during 13 solar cycles can be interpreted as evidence of
the rigid rotation of active longitudes.

If we use for the phase of solar cycle the following notation: $k = 1$ corresponds to the period of 11-year cycle from the minimum to the polar
magnetic field reversal; $k = 2$ corresponds to the period from the reversal to the minimum, then the location of the distribution maximum (the
active longitude) can be represented by the formula:
\begin{equation}
\label{2}
 Active\,\, Longitude = \pi k
\end{equation}

From the solar activity minimum to the reversal of the polar field (the AM  period, active longitude $180^\circ$) the sign of the polar magnetic
field and the sign of leading sunspots in a hemisphere coincide, while for the period from the reversal to the solar minimum (the DM period,
active longitude $\sim 0^\circ/360^\circ$) these polarities are opposite.


\section{Conclusions}
\label{S-Conclusions}

Introducing the notion of the longitudinal asymmetry vector we evaluated the nonaxisymmetric part of the solar activity distribution. This
approach allowed to reduce the influence of the stochastic component of the solar activity uniformly distributed over longitude and separate the
stable nonaxisymmetric component. The change in the longitudinal asymmetry of the sunspot distribution was analyzed for the period of 13 solar
cycles (1874--2016). The study of the nonaxisymmetric component of solar activity showed, on the one hand, a close connection of this component
with the development of the solar cycle. On the other hand, significant features of longitudinal asymmetry have been found. In the longitudinal
asymmetry of sunspot distribution for the disc and for the southern hemisphere in each four subsequent cycles (from Solar Cycle 12 to Solar
Cycle 23) only for the first pair of cycles the Gnevyshev--Ohl  rule held, while the next pair obeyed the `anti-rule', i.e., the amplitude of an
even cycle was higher than the amplitude of the next odd one. Possibly, this is a manifestation of a 44-year structure of solar activity.

For the northern hemisphere the Gnevyshev--Ohl rule holds for Solar Cycles 12--17, while for Solar Cycles 18--23 the anti-rule can be observed.
The longitudinal asymmetry for sunspots with area $Sp < 50$ MSH follows the Gnevyshev--Ohl rule similar to the sunspot area. Thus, the
alternation of the Gnevyshev--Ohl rule and anti-Gnevyshev--Ohl rule is characteristic for large long-living sunspots.

The longitudinal asymmetry integral obeys the pattern found in (\opencite{gnev48}): the high correlation of the even $(2N)$ and the subsequent
odd cycle $(2N+1)$ is accompanied by the low correlation of the even cycle $(2N)$ and preceding odd cycle $(2N-1)$.

The longitudinal distribution of the sunspot area for the period of  13 solar cycles does not show the presence of preferred longitudes. The
concentration of solar activity on certain longitudes can be observed when the phase of the longitudinal asymmetry vector is considered for the
periods ascent-maximum and descent-minimum separately. Then it becomes evident that the phase of the vector has the maximum at the longitude
$180^\circ$ during the ascent-maximum epoch and at the longitude $0^\circ/360^\circ$ during the descent-minimum epoch. A drastic change of the
active longitude by $180^\circ$ (flip-flop) occurs near the solar activity minimum and at the Sun's global field reversal.  The active longitude
changes with the reversals of polarity of the local and global magnetic fields.

%
 \begin{acks}
  We thank the staff of the RGO/USAF/NOAA and personally Dr.~David H. Hathaway for providing the homogeneous set of the sunspot data.
 \end{acks}
\medskip
\\
\noindent {\bf Disclosure of Potential Conflicts of Interest:} The
authors declare that they have no conflicts of interest


\end{article}

\end{document}